\documentclass{article}
\usepackage{graphicx}
\begin{document}
\title{ Machine Learning of Nonlinear Dynamical Systems with Control Parameters Using Feedforward Neural Networks}
\author{Hidetsugu Sakaguchi\\
Interdisciplinary Graduate School of Engineering Sciences, \\
Kyushu
University, Kasuga, Fukuoka 816-8580, Japan}
\maketitle
\begin{abstract}
Several authors have reported that the echo state network reproduces bifurcation diagrams of some nonlinear differential equations using the data for a few control parameters. We demonstrate that a simpler feedforward neural network can also reproduce the bifurcation diagram of the logistics map and synchronization transition in globally coupled Stuart-Landau equations.  
\end{abstract}
Deep learning using multi-layer neural networks is an important tool in artificial intelligence~\cite{Goodfellow}. The learning cost is high when the number of layers and neurons is large. The echo state network (ESN) has a three-layer structure consisting of the input, intermediate, and output layers~\cite{Jaeger}.  The neurons in the intermediate layer interact with each other.  Only the connection strength between the intermediate and output layers changes during the learning process. The ESN is applied to learn the time series. Recently, several authors have reported that bifurcation phenomena in nonlinear differential equations can be reproduced from data for a few control parameters using the ESN~\cite{Fan}. In the ESN, mutual interaction in the intermediate layer makes the dynamics of the network versatile and the memory of previous inputs is effectively stored in the intermediate layer. If mutual interactions inside the intermediate layer are absent, then the neural network becomes a simple feedforward network. The output is uniquely determined by the input. In this study, we investigate whether the bifurcation phenomena can be reproduced using the data for a few control parameters even in the simple feedforward network.

Feedforward networks are constructed with three layers: input, intermediate, and output layers. The input and intermediate layers are randomly connected with the connection strength $C_{j,i}$. The response function for each neuron is assumed to be $y=\tanh x$. Then, the response $y_j$ of the $j$th intermediate neuron is expressed as   
\begin{equation}
y_j=\tanh(\sum_{i=1}^{Q} C_{j,i}x_i-h_j) \;\;\;{\rm for}\;\;j=1,2,\cdots,N,
\end{equation}
where $x_i$ is the $i$th input value, and $h_j$ is the threshold. $Q$ and $N$ denote the numbers of neurons in the input and intermediate layers, respectively. The connection strength $C_{j,i}$ and threshold $h_j$ are randomly selected from a uniform random number, and are fixed. 
The response of the $k$th neuron in the output layer is determined by the linear function of $y_j$ in the intermediate layer as follows:
\begin{equation}
z_k=\sum_{j=1}^ND_{k,j}y_j\;\;\;{\rm for}\;\;k=1,2,\cdots,P,
\end{equation}
where $P$ denotes the number of neurons in the output layer. 
The connection strength $D_{k,j}$ is determined using the Ridge regression. 
Rosenblatt's perceptron has the same three-layer feedforward network, but the output is expressed as $\theta(z_k-h_k^{\prime})$  where $\theta(x)$ is the Heaviside step function~\cite{Rosenblatt}, and the error-correction algorithm is used for learning. In the Ridge regression, the matrix ${\bf Y}$ is defined as follows:
\[
{\bf Y}=[{\bf y}(1),{\bf y}(2),\cdots, {\bf y}(m)]
\]
where the vector ${\bf y}(n)$ at timestep $n$ is the column vector $(y_1(n),y_2(n),\cdots,y_N(n))^T$ for the response of $N$ neurons in the intermediate layer (${\bf v}^T$ denotes the transpose of vector ${\bf v}$), and a matrix ${\bf Z}$ is defined as
\[{\bf Z}=[{\bf z}(1),{\bf z}(2),\cdots, {\bf z}(m)]
\]
where ${\bf z}(n)$ is the column vector $(z_1(n),z_2(n),\cdots, z_P(n))^T$ of the target values of the $P$ outputs at timestep $n$.    
$D_{k,j}$ is calculated as
\begin{equation}
{\bf D}={\bf Z}{\bf Y}^T({\bf Y}{\bf Y}^T+\beta {\bf I})^{-1}
\end{equation}
where $\beta$ is a regularization parameter and ${\bf I}$ is the unit matrix. In this short note, we present two examples, a logistic map and coupled Stuart-Landau oscillators to demonstrate the validity of the feedforward network. 

First, we demonstrate the machine learning of the logistic map. The logistic map is a typical map that depicts the chaos. The logistic map is expressed as 
\begin{equation}
x_{n+1}=ax_n(1-x_n),
\end{equation}
where $0<a<4$ denotes the control parameter. We demonstrate that a bifurcation diagram can be reproduced using a feedforward neural network. We assume that $x_{n}$ and $a$ are inputs and $x_{n+1}$ is the output. The output $x_{n+1}$ is considered as the input for the next timestep. That is, feedback from the output layer to the input layer with a time delay of 1 is assumed. The number of neurons in the intermediate layer is $N=500$, $C_{j,i}$ is a random number between -1 and 1, and $h_j$ is a random number between -0.5 and 0.5. The parameter $\beta$ is set to 0.01. The data of $x_n$ for $n=1,2,\cdots, 100$ for $a=2.8$, 3.4, and 3.9 were used for the Ridge regression. After learning, the outputs $x_n$ for $a$ different from 2.8, 3.4, and 3.9 were calculated using the feedforward network. 

Figure 1(a) shows the relationship between the input $x_n$ and output $x_{n+1}$ for $n=1,2,\cdots 500$ obtained by the feedforward network for the parameter $a=3.8$. The dashed line represents $y=ax(1-x)$ for $a=3.8$. The input-output relationship of the logistic map  was well reproduced by the feedforward network. 
Figure 1(b) shows the bifurcation diagram obtained by varying $a$. Period-doubling bifurcation and chaos were well reproduced in the feedforward network, although the bifurcation points  deviated slightly from those of the logistic map.     
\begin{figure}[h]
\begin{center}
\includegraphics[height=3.8cm]{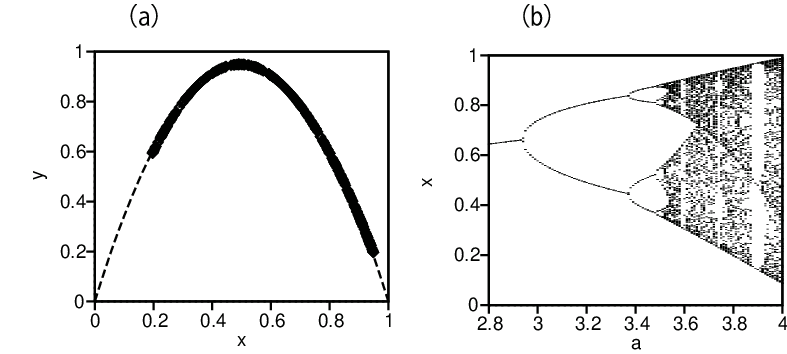}
\end{center}
\caption{(a) The input-output relationship at $a=3.8$. The dashed line represents $y=3.8x(1-x)$. (b) Bifurcation diagram between $2.8\le a\le 4$ by the feedforward network after the learning process of the Ridge regression using the data of the logistic map.}
\label{f1}
\end{figure}

Next, we apply the method to mutual synchronization in globally coupled Stuart-Landau equations. As a model of mutual synchronization, we consider the coupled equations of the Stuart-Landau model as follows:
\begin{equation}
\frac{dW_i}{dt}=(1+\omega_i)W_i-|W_i|^2W_i+K(\frac{1}{M}\sum_{j=1}^MW_j-W_i),
\end{equation}
where $W_i=X_i+iY_i$ ($i=1,2,\cdots,M$) is a complex variable, the total number $M$ of oscillators is set to $M=1000$, and the natural frequency $\omega_i$ is selected from a Gaussian random number with an average of 0 and a standard deviation of 0.05. The solid line in Fig.~2(a) shows the order parameter $S=|\sum_{j=1}^M W_j/M|$ as a function of coupling strength $K$. A phase transition occurs owing to mutual synchronization near $K=0.075$. 

We consider a three-layer neural network to approximate the nonlinear dynamics of each oscillator. M copies of the three-layer networks are used for the coupled system of $M$ oscillators. The input data are  $X_i(t)$, $Y_i(t)$, $\omega_i$, $K$, $\sum X_i/M$, and $\sum Y_i/M$, and the outputs are $X_i(t+\tau)$, $Y_i(t+\tau)$ with $\tau=0.2$.
In the learning process, we used 10000 data points of $X_i(n\tau)$, $Y_i(n\tau)$ obtained by direct numerical simulation of the coupled Stuart-Landau equations of three oscillators ($M=3$) with $\omega_i=0.05,0.02$, and -0.03. $K=0.05$, 0.1, and 0.15 were selected as the coupling strengths for learning. The number of neurons $N$ in the intermediate layer was set to $N=575$, and $\beta=0.01$. $C_{j,i}$ is a random number between -1 and 1, and $h_j$ is a random number between -0.5 and 0.5. $C_{j,i}$, $h_j$, and  $D_{k,j}$ are assumed to have the same values for the three networks. Figure 2(b) shows  $X_i(t)$ in the coupled Stuart-Landau equation and the feedforward network for $\omega_i=-0.03$ and $K=0.05$, where the same initial values are used. The nonlinear dynamics of the coupled Stuart-Landau equations is well reproduced by the feedforward network. 

The dashed line in Fig.~2(a) shows the order parameter $S$ as a function of $K$ obtained by the feedforward network of $M=1000$, where $C_{j,i}$ and $D_{k,j}$ are the same as those obtained by the learning process of $M=3$, and $\omega_i$ takes the same Gaussian random number as that in Eq.~(5). We input the same values $\sum X_i/M$, and $\sum Y_i/M$ into each network. The parameters $K$ and $\omega_i$ are fixed in time, but $X_i$, $Y_i$, $\sum X_i/M$, and $\sum Y_i/M$ change over time. The time-averaged value of the order parameter is plotted as the dashed line in Fig.~2(a). Although the order parameter is slightly different from that of the solid line, the phase transition is reproduced well in the feedforward network.     
\begin{figure}[h]
\begin{center}
\includegraphics[height=3.8cm]{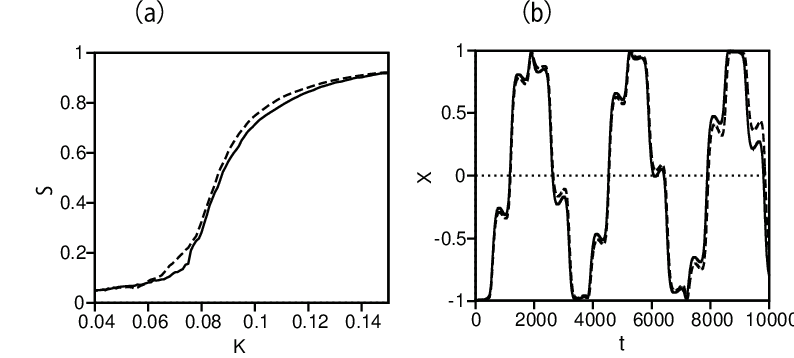}
\end{center}
\caption{(a) Order parameter $S$ (solid line) in the coupled equations of the Stuart-Landau oscillators with $M=1000$ as a function of $K$ and the order parameter (dashed line) obtained by the feedforward neural network. (b) Time sequence (solid line) of $X_i(t)$ in the coupled Stuart-Landau equations of $M=3$ and a time sequence of $X_i(t)$ by the three-layer feedforward network with $N=575$ neurons in the intermediate layer (dashed line). }
\label{f1}
\end{figure}

In summary, we demonstrated that a simple three-layer feedforward network can reproduce the bifurcation diagram of the logistic map and synchronization transition in globally coupled Stuart-Landau equations using data for a few control parameters. We showed that nonlinear mapping close to $ax(1-x)$ is approximately reproduced in the feedforward network for the logistic map. That is related to the universal approximation theorem in which an arbitrary function can be approximated in a feedforward network~\cite{Kurt}. The advantage of the feedforward network is that the computation time is reduced compared to that of the echo state network because we do not need to repeat the multiple interation of $N\times N$ calculations in the intermediate layer.

\end{document}